\documentclass[aps,twocolumn,a4paper,10pt]{revtex4-1}
\usepackage{mathrsfs}
\usepackage{graphicx}
\usepackage{latexsym}
\usepackage{amsmath}
\usepackage{amssymb}
\usepackage{textcomp}
\usepackage{amsbsy}
\usepackage{epstopdf}

\begin{document}
 \title{\large \bf Radion stabilization in higher curvature warped spacetime}
 \author{Ashmita Das}
\email{ashmita@iitg.ernet.in}
\affiliation{Department of Physics, Indian Institute of Technology, North Guwahati, Guwahati, Assam 781039, India}
\author{Hiya Mukherjee}
\email{tphm@iacs.res.in}
\author{Tanmoy Paul }
\email{tpap@iacs.res.in}
\author{Soumitra SenGupta}
\email{tpssg@iacs.res.in}
\affiliation{Department of Theoretical Physics,\\
Indian Association for the Cultivation of Science,\\
2A $\&$ 2B Raja S.C. Mullick Road,\\
Kolkata - 700 032, India.\\}

\begin{abstract}
We consider a five dimensional AdS spacetime in presence of higher curvature term like $F(R) = R + \alpha R^2$ 
in the bulk. 
In this model, we examine the possibility of modulus stabilization 
from the scalar degrees of freedom of higher curvature gravity free of ghosts.
Our result reveals that the model stabilizes itself and the mechanism of modulus stabilization 
can be argued from a geometric point of view. We 
determine the region of the parametric space for which the modulus (or radion) can to be stabilized. 
We also show how the mass and coupling parameters 
of radion field are modified due to higher curvature term leading to modifications 
of its phenomenological implications on the visible 3-brane.
\end{abstract}
\maketitle

\section{Introduction}
Till date, Standard Model (SM) of particle Physics is a widely accepted theory 
to describe the interactions of fundamental particles. Despite its enormous 
successes, the model is plagued with divergence of the Higgs mass due to radiative 
corrections which may run up to Planck scale. An unnatural fine tuning is needed to confine 
the Higgs mass within TeV scale.

Many attempts have been made to address this problem by considering 
the theories beyond SM of particle Physics. Few such candidates are - 
supersymmetry, technicolor and extra dimensions. Among many such attempts \cite{arkani,horava,RS1,kaloper,cohen,burgess,chodos}, 
Randall-Sundrum (RS) model \cite{RS1} of warped extra dimension draws special attention 
since it resolves the gauge hierarchy problem without choosing  
 any intermediate scale in the theory.

In RS model, two 3-branes are embedded in a five dimensional anti de-Sitter spacetime 
compactified on a $M_4 \times S_1/Z_2$ orbifold. The distance between these two branes 
is assumed to be $\sim$ of Planck length so that the required 
hierarchy between the two branes is generated.
One of the crucial aspects of this braneworld scenario is to stabilize 
the distance between the branes (known as modulus or radion). For this 
purpose, it is necessary to generate an appropriate radion potential 
with a stable minimum consistent with the value proposed in RS model 
in order to solve the hierarchy problem. Goldberger and Wise (GW) proposed 
a mechanism \cite{GW1} to create such a radion potential by introducing a 
bulk scalar field with appropriate boundary values. Subsequently the phenomenology 
of radion field has also been studied extensively. This radion phenomenology \cite{GW_radion,kribs,julien,wolfe} 
along with the study of RS graviton \cite{dhr,rizzo,yong,dhr1,thomas} and RS black holes 
\cite{wiseman,naresh,dai} are considered to be the testing ground of warped extra dimensional models in collider experiments \cite{atlas1,atlas2}.
As the present experimental lower bound of the  first graviton KK-mode mass
climbs above 3 Tev, the RS-type resolution of the naturalness problem is undoubtedly 
under pressure. The question that how the Higgs is so much lighter than the 5-dimensional 
Planck scale needs to be settled properly. However, in any higher dimensional model with 
gravity in the bulk, the  modulus must be stabilized to appropriate value to extract a meaningful
low energy effective theory on the brane. The present work aims to address this issue specially
in the context of a higher dimensional model where the fundamental curvature scale is 
of the order of Planck scale.\\

It is well known that Einstein-Hilbert action can be generalized
by adding higher order curvature terms 
which naturally arise from diffeomorphism property of the action. 
Such terms also have their origin in String theory due to 
quantum corrections. 
$F(R)$ \cite{faraoni,felice,paliathanasis}, Gauss-Bonnet (GB) \cite{nojiri2,nojiri3,cognola}
or more generally Lanczos-Lovelock gravity 
are some of the candidates in higher curvature gravitational 
theory.

Higher curvature terms become extremely relevant at the regime 
of large curvature. Thus for RS bulk geometry where the curvature is of the  
order of Planck scale, the higher curvature terms should play 
a crucial role.  In general inclusion of higher curvature terms in the action leads to the appearance of 
ghost from higher derivative  terms resulting into Ostragradsky instability. 
The Gauss-Bonnet model ( a special case of Lanczos-Lovelock  model ) is however free of this instability due to appropriate choice of various quadratic combinations of Riemann tensor, Ricci tensor and curvature scalar . Some important  modified solutions of the Randall-Sundrum model in presence of Gauss-Bonnet terms have been obtained by Kim et.al  \cite{kim1, kim2}  in the context of both static and inflationary scenario. A GB modified warped solution and it's phenomenological implications was also discussed in \cite{ssgsc}.

In contrast to GB model $F(R)$ gravity model however contains higher curvature terms consisting only of the scalar curvature $R$. Once again just as GB model, certain classes of $F(R)$ gravity models are free from ghost-like instability ( See section III  for a detailed discussion ).
 In general  $F(R)$ model can be mapped into a scalar-tensor 
theory at the action level by a conformal transformation of the metric 
\cite{faraoni,felice,barrow,marino,bahamonde,catena,ssg1,ssg2} . 
The issue of instability of the original $F(R)$ model is now reflected in the form of the kinetic and potential terms of the scalar field in the dual scalar-tensor model, where the potential  will have a stable minimum and a kinetic term with proper signature  if the original $F(R)$ model is free from  ghosts ( see section III ).
 It is know that to achieve modulus stabilization  in RS-like model,  one needs a scalar field \cite{GW1}. We, in this work, therefore look for  the possibility of having a geometric  modulus stabilization mechanism due to the scalar degree of freedom originating from the higher curvature terms. 
The important questions in this context are :
\begin{itemize}
 \item Is the RS braneworld modified by $F(R)$ gravity,  
 stabilized even without introducing an external stabilizing field ?
 
 \item If the  modulus can be  stabilized in the dual scalar-tensor model,  does it mean that it is also  stabilized in the original $F(R)$ model ?
 
 \item Does the scalar kinetic and potential terms for the purpose of modulus stabilization correspond to a $F(R)$ model which is free of ghost-like instability ?
 
 \item If the braneworld scenario is 
 stabilized consistently  then how the radion mass and coupling parameters will 
 change from that of RS scenario due to 
 the presence of higher order curvature terms ?
\end{itemize}
We aim to address these questions in this work by using the 
equivalence between $F(R)$ and scalar-tensor theory.

The paper is organized as follows: Following two sections are devoted to  
brief reviews of RS scenario and conformal relationship between $F(R)$ 
and scalar-tensor (ST) theory. In section IV, we extend our analysis 
of section III for the specific $F(R)$ model considered in this work. 
Section V extensively describes the modulus stabilization, radion mass 
and coupling for the dual ST model while section VI addresses these for the 
original $F(R)$ model. After discussing the equivalence, the paper 
ends with some conclusive remarks in section VII.

\section{Brief description of RS scenario and its stabilization via GW mechanism}

RS scenario is defined on a five dimensional AdS spacetime involving one warped and compact 
extra spacelike dimension. Two 3-branes known as TeV/visible and Planck/hidden brane 
are embedded in five dimensional spacetime where the intermediate region 
between the branes is termed as 'bulk'. If $\phi$ is the extra dimensional angular 
coordinate, then the branes are 
located at two fixed points $\phi=(0,\pi)$ while the latter one is identified 
with our known four dimensional universe. The opposite brane tensions along with the finely 
tuned five dimensional cosmological constant serve as energy-momentum tensor of RS 
scenario. The resulting spacetime metric \cite{RS1} is non-factorizable and expressed as,
\begin{equation}
ds^2 = e^{- 2 kr_c|\phi|} \eta_{\mu\nu} dx^{\mu} dx^{\nu} -r_c^2d\phi^2 \label{eq1}
\end{equation}
Here, $r_c$ is the compactification radius of the extra dimension. 
Due to $S^1/Z_2$ compactification along the extra dimension, $\phi$ ranges from 
$-\pi$ to $+\pi$.
The quantity $k=\sqrt{\frac{-\Lambda}{12M^3}}$, is of the order of 5-dimensional Planck
scale $M$. Thus $k$ relates the 5D Planck scale $M$ to the 5D cosmological constant
$\Lambda$.\\
All the dimensionful parameters described above are
related to the reduced 4-dimensional Planck scale ${M}_{Pl}$ as,
\begin{equation}
 M_{Pl}^2=\frac{M^3}{k}(1-e^{-2k\pi r_c})\label{rplanckmass}
\end{equation}

In order to solve the hierarchy problem, it is assumed in RS scenario that the branes 
are separated by such a distance that 
$k\pi r_c \approx 36$. Then the exponential factor present
in the metric, which is often called warp factor, produces 
a large suppression so that a mass scale of the order of Planck scale is reduced to TeV scale on the 
visible brane. 
A scalar mass say mass of Higgs is given as, 
\begin{equation}
 m_H=m_{0}e^{-k\pi r_c}\label{physmass}
\end{equation}
where $m_H$ and $m_0$ are physical and bare Higgs mass respectively. 
But one of the crucial aspects of this braneworld scenario is to stabilize 
the distance between the branes (known as modulus or radion).
For this purpose, Goldberger and Wise demonstrated that the modulus corresponding to the
 radius of the extra dimension in RS warped geometry model can be stabilized \cite{GW1} by invoking a massive 
scalar field in the bulk with appropriate vacuum expectation values
(vev) at the two 3-branes that reside at the orbifold fixed points. 
Consequently the phenomenology of the radion 
field originating from 5D gravitational degrees of freedom has also been explored \cite{GW_radion}.

\section{Transformation of a F(R) theory to scalar-tensor theory}

In this section, we briefly describe how a higher curvature F(R) gravity model in five dimensional 
scenario can be recast into Einstein gravity with a scalar field. 
The F(R) action is expressed as,
\begin{equation}
 S = -\frac{1}{2\kappa^2}\int d^4x d\phi \sqrt{G} F(R)
 \label{action0}
\end{equation}
where $x^{\mu} = (x^0, x^1, x^2, x^3)$ are usual four dimensional coordinate and 
$\phi$ is the extra dimensional spatial angular coordinate. 
$R$ is the five dimensional Ricci curvature and $G$ is the determinant of the metric. 
Moreover $\frac{1}{2\kappa^2}$ as taken as $2M^3$ where $M$ is the five dimensional 
Planck scale. Introducing an auxiliary field $A(x,\phi)$, 
above action (\ref{action0}) can be equivalently written as,
\begin{equation}
 S = -\frac{1}{2\kappa^2}\int d^4x d\phi \sqrt{G} [F'(A)(R-A) + F(A)]
 \label{action00}
\end{equation}
By the variation of the auxiliary field $A(x,\phi)$, one easily obtains $A=R$. 
Plugging back this solution $A=R$ into action (\ref{action00}), 
initial action (\ref{action0}) can be reproduced. At this stage, perform a 
conformal transformation of the metric as
\begin{equation}
  G_{MN}(x,\phi) \rightarrow \tilde{G}_{MN} = \exp{(\sigma(x,\phi)}G_{MN}(x,\phi)
 \nonumber\\
\end{equation}
$M, N$ run form 0 to 5. $\sigma(x,\phi)$ is conformal factor and related to the auxiliary 
field as $\sigma = (2/3)\ln F'(A)$. If $R$ and $\tilde{R}$ are the Ricci scalar formed by 
$G_{MN}$ and $\tilde{G}_{MN}$ respectively, then 
they are related as,
\begin{eqnarray}
 R = e^{\sigma} \bigg[\tilde{R} - 4\tilde{G}^{MN} \partial_M\partial_N\sigma 
 - 3 \tilde{G}^{MN} \partial_M\sigma\partial_N\sigma\bigg]
 \nonumber
\end{eqnarray} 

Due to the above relation between $R$ and $\tilde{R}$, action (\ref{action00}) turns out to be,
\begin{eqnarray}
 S&=&\int d^4x d\phi \sqrt{\tilde{G}} \bigg[-\frac{1}{2\kappa^2}e^{-\frac{3\sigma}{2}}F'(A)\bigg(\tilde{R} 
 -4\tilde{G}^{MN} \partial_M\partial_N\sigma\nonumber\\ 
 &-&3\tilde{G}^{MN}\partial_M\sigma \partial_N\sigma\bigg) - \frac{1}{2\kappa^2}e^{-\frac{5\sigma}{2}}\bigg(F(A) - AF'(A)\bigg)\bigg]
 \label{new}
\end{eqnarray}

Considering $F'(R) > 0$ and using the aforementioned relation between $\sigma$ and $F'(A)$, 
one lands up to the following scalar-tensor action
\begin{eqnarray}
 S=(1/2\kappa^2)\int d^4x d\phi \sqrt{\tilde{G}} \bigg[-\tilde{R} + 3\tilde{G}^{MN}\partial_M\sigma \partial_N\sigma \nonumber\\
 - \frac{1}{2\kappa^2}\bigg(\frac{F(A)}{F'(A)^{5/3}} - \frac{A}{F'(A)^{2/3}}\bigg)\bigg]
 \label{new1}
\end{eqnarray}
where $\tilde{R}$ is the Ricci scalar formed by $\tilde{G}_{MN}$. $\sigma(x,\phi)$ is 
the scalar field, emerging from 
higher curvature degrees of freedom. At this stage, it is important to note  that for $F'(R) < 0$, the kinetic term of the scalar field 
$\sigma$ as well as the Ricci scalar $\tilde{R}$ in the above action come with wrong sign, which indicates the existence of ghost field. Thus to avoid 
the ghost like structure, $F'(R)$ must be greater than zero. Later we shall show that in the context of the present work, this condition is indeed satisfied.\\ 
The kinetic part of $\sigma(x,\phi)$ in eqn.(\ref{new1}), though correct in sign, is
non-canonical. In order to make the scalar field canonical, transform 
$\sigma$ $\rightarrow$ $\Phi(x,\phi) = \sqrt{3}\frac{\sigma(x,\phi)}{\kappa}$. In terms of 
$\Phi(x,\phi)$, the above action takes 
the form,
\begin{equation}
 S=\int d^4x d\phi \sqrt{\tilde{G}} \bigg[-\frac{\tilde{R}}{2\kappa^2} + \frac{1}{2}\tilde{G}^{MN}\partial_M\Phi 
 \partial_N\Phi - V(\Phi)\bigg]
 \nonumber\\
\end{equation}

where $V(\Phi) = \frac{1}{2\kappa^2}[\frac{F(A)}{F'(A)^{5/3}} - \frac{A}{F'(A)^{2/3}}]$ is 
the scalar field potential which depends on the 
form of $F(R)$. Thus the action of $F(R)$ gravity in five dimension can be transformed 
into the action of a scalar-tensor 
theory by a conformal transformation of the metric.

\section{RS like spacetime in F(R) model and corresponding scalar-tensor theory}

In the present work, we consider a five dimensional AdS spacetime with two 3-brane scenario in F(R) model. To the leading order in higher curvature term, the form of 
$F(R)$ is taken as $F(R) = R + \alpha R^2$ where $\alpha$ is a constant with square of the 
inverse mass dimension. Considering $\phi$ as the extra dimensional 
angular coordinate, two branes are located at $\phi = 0$ (hidden brane) and at $\phi = \pi$ (visible brane) respectively 
while the latter one is identified with the visible universe. Moreover the extra dimension  is $S^1/Z_2$ 
orbifolded along the coordinate $\phi$. 
The action for this model is :
\begin{eqnarray}
 S&=&\int d^4x d\phi \sqrt{G} \bigg[-\frac{1}{2\kappa^2}(R + \alpha R^2) - \Lambda\nonumber\\ 
 &-&V_h\delta(\phi) - V_v\delta(\phi-\pi)\bigg]
 \label{actionF(R)}
\end{eqnarray}
where $\Lambda (< 0)$ is the bulk cosmological constant and $V_h$, $V_v$ are the brane tensions on hidden, visible brane 
respectively.

This higher curvature like $F(R)$ model (in eqn.(\ref{actionF(R)})) can be transformed into scalar-tensor 
theory by using the technique discussed in the previous section. Performing a conformal transformation of the metric as 
\begin{equation}
 G_{MN}(x,\phi) \rightarrow \tilde{G}_{MN} = \exp{(\frac{1}{\sqrt{3}}\kappa\Phi(x,\phi))}G_{MN}(x,\phi)
 \label{conformal}
\end{equation}
the above action (in eqn.(\ref{actionF(R)})) can be expressed as a scalar-tensor theory with the action given by :
\begin{eqnarray}
 S&=&\int d^4x d\phi \sqrt{\tilde{G}} \bigg[-\frac{\tilde{R}}{2\kappa^2} + \frac{1}{2}\tilde{G}^{MN}\partial_M\Phi \partial_N\Phi - V(\Phi)\nonumber\\
 &-&\Lambda - \exp{(-\frac{5}{2\sqrt{3}}\kappa\Phi)} V_h\delta(\phi)\nonumber\\ 
 &-&\exp{(-\frac{5}{2\sqrt{3}}\kappa\Phi)} V_v\delta(\phi-\pi)\bigg]
 \label{action1ST}
\end{eqnarray}
where the quantities in tilde are reserved for ST theory. $\tilde{R}$ is the Ricci curvature formed 
by the transformed metric $\tilde{G}_{MN}$. $\Phi(x,\phi)$ is the scalar 
field corresponds to higher curvature degrees of freedom and $V(\Phi)$ is the scalar potential which 
for this specific choice form of $F(R)$ has the form,
\begin{eqnarray}
 V(\Phi)&=&-\frac{1}{8\kappa^2\alpha} \exp{(-\frac{5}{2\sqrt{3}}\kappa\Phi)}[\exp{(\frac{3}{2\sqrt{3}}\kappa\Phi)} - 1]^2\nonumber\\
 &+&\Lambda [\exp{(-\frac{5}{2\sqrt{3}}\kappa\Phi)}-1]
 \label{scalar_potential}
\end{eqnarray}

One can check that the above potential (in eqn.(\ref{scalar_potential})) is stable for the 
parametric regime $\alpha < 0$. This immediately ensures that for our model with negative bulk curvature, $F'(R)  = 1 + \alpha R $  is always greater than zero which in turn indicates that the original $F(R)$ model is free from Ostragradsky instability as discussed earlier.

The stable value ($<\Phi>$) as well as the mass squared ($m_{\Phi}^2$) 
of the scalar field ($\Phi$) are given by the following two equations
\begin{equation}
 \exp{(\frac{3}{2\sqrt{3}}\kappa<\Phi>)} = [\sqrt{9 - 40\kappa^2\alpha\Lambda} - 2]
 \label{vev_phi}
\end{equation}
and
\begin{equation}
 m_{\Phi}^2 = -\frac{1}{8\alpha} [\sqrt{9 - 40\kappa^2\alpha\Lambda}] [\sqrt{9 - 40\kappa^2\alpha\Lambda} - 2]^{-\frac{2}{3}}
 \label{mass_phi}
\end{equation}
Furthermore, the minimum value of the potential i.e. $V(<\Phi>)$ is non zero and serves as a cosmological constant. Thus 
the effective cosmological constant in scalar-tensor theory is $\Lambda_{eff} = \Lambda + V(<\Phi>)$ where $V(<\Phi>)$ 
is,
\begin{eqnarray}
 V(<\Phi>)&=&-\Lambda + [\sqrt{9 - 40\kappa^2\alpha\Lambda} - 2]^{-\frac{5}{3}}\nonumber\\
 &[&\Lambda - (1/8\kappa^2\alpha)[\sqrt{9 - 40\kappa^2\alpha\Lambda}-3]^2]
 \nonumber\\
\end{eqnarray}
Above form of $V(<\Phi>)$ with $\Lambda < 0$ clearly indicates that $\Lambda_{eff}$ is also 
negative or more explicitly,the corresponding scalar-tensor theory for the original $F(R)$ 
model has an AdS like spacetime. Considering $\xi$ as the fluctuation 
of the scalar field over its vev, the final form of action for the scalar-tensor theory in the bulk can be written as,
\begin{eqnarray}
 S&=&\int d^4x d\phi \sqrt{\tilde{G}} \bigg[-\frac{\tilde{R}}{2\kappa^2} + \frac{1}{2}\tilde{G}^{MN}\partial_M\xi 
 \partial_N\xi\nonumber\\
 &-&(1/2)m_{\Phi}^2\xi^2 - \Lambda_{eff}]
 \label{action2ST}
\end{eqnarray}
where the terms up to quadratic order in $\xi$ are retained for $\kappa\xi < 1$. A detailed 
justification of neglecting the higher order terms as well as their possible effects 
will be discussed in section VI-C.

\section{Modulus stabilization, radion mass and coupling in scalar-tensor (ST) theory}

\subsection{Modulus stabilization}
In order to stabilize the modulus in ST theory, here we adopt the GW mechanism \cite{GW1} which requires 
a massive scalar field in the bulk. For the case of ST theory presented in eqn.(\ref{action2ST}), $\xi$ can act as a bulk 
scalar field with the mass given by eqn.(\ref{mass_phi}). Considering a negligible backreaction 
of the scalar field ($\xi$) on the background spacetime, the solution of metric $\tilde{G}_{MN}$ 
is exactly same as RS model i.e.
\begin{equation}
 d\tilde{s}^2 = e^{- 2 kr_c|\phi|} \eta_{\mu\nu} dx^{\mu} dx^{\nu} - r_c^2d\phi^2
 \label{grav.sol1.ST}
\end{equation}
where $k = \sqrt{\frac{-\Lambda_{eff}}{24M^3}}$. With this metric, the scalar field equation of motion 
in the bulk is following,
\begin{eqnarray}
 &-&\frac{1}{r_c^2}\partial_\phi[\exp{(-4kr_c|\phi|)}\partial_\phi\xi]\nonumber\\
 &+&m_{\Phi}^2\exp{(-4kr_c|\phi|)}\xi(\phi) = 0
 \label{eom.scalar.field}
\end{eqnarray}
where the scalar field $\xi$ is taken as function of extra dimensional coordinate only. Considering 
non zero value of $\xi$ on branes, the above equation (\ref{eom.scalar.field}) has the general 
solution,
\begin{equation}
 \xi(\phi) = e^{2kr_c|\phi|} \big[Ae^{\nu kr_c|\phi|} + Be^{-\nu kr_c|\phi|}\big]
 \label{sol.scalar.field}
\end{equation}
with $\nu = \sqrt{4 + m_{\Phi}^2/k^2}$. Moreover $A$ and $B$ are obtained from the 
boundary conditions, $\xi(0)=v_h$ and $\xi(\pi)=v_v$ as follows :
\begin{equation}
 A = v_v e^{-(2+\nu)kr_c\pi} - v_h e^{-2\nu kr_c\pi}
 \label{A}
\end{equation}
and
\begin{equation}
 B = v_h (1 + e^{-2\nu kr_c\pi}) - v_v e^{-(2+\nu)kr_c\pi}
 \label{B}
\end{equation}
Upon substitution the form of $A$ and $B$ into eqn.(\ref{sol.scalar.field}), one finds 
that 
\begin{eqnarray}
\xi(0)&=&{v}_h\nonumber\\
\xi(\pi)&=&v_v[1-e^{-2\nu kr_c\pi}+\frac{v_h}{v_v}e^{-(3\nu-2)kr_c\pi}]
\nonumber
\end{eqnarray}
Above values of $\xi(0)$ and $\xi(\pi)$ matches with the boundary condition (i.e. $\xi(0)=v_h$ and $\xi(\pi)=v_v$) 
by neglecting the subleading powers of $e^{-kr_c\pi}$, as have been done earlier 
by the authors in \cite{GW1}.\\
Plugging back the solution of $\xi(\phi)$ (eqn.(\ref{sol.scalar.field})) into scalar field action and 
integrating over $\phi$ yields an effective modulus potential having the following form,
\begin{equation}
 V_{eff}(r_c) = k(\nu + 2)A^2(e^{2\nu kr_c\pi} - 1) + k(\nu - 2)B^2(1 - e^{-2\nu kr_c\pi})
 \label{mod.potential}
\end{equation}
This potential has a minimum at
\begin{equation}
 k\pi r_c = \frac{4k^2}{m_{\Phi}^2}\ln{[\frac{v_h}{v_v}]}
 \label{stabilized.modulus.ST}
\end{equation}
According to GW mechanism \cite{GW1}, the branes are stabilized at that separation for which the 
modulus potential becomes minimum. Thus the above equation represents the stabilized value 
for inter-brane separation. Expression of $m_{\Phi}^2$ (eqn.(\ref{mass_phi})) clearly indicates 
that $r_c$ in eqn.(\ref{stabilized.modulus.ST}) 
is positive only for $\alpha < 0$. Thus the $F(R)$ model depicted in eqn.(\ref{actionF(R)}) with positive $\alpha$ 
can be transformed to a scalar-tensor theory where the scalar field has negative squared mass and 
the modulus of the ST theory can not be stabilized.\\
It may be observed that  the scalar field degrees of 
freedom is related to the curvature as,
\begin{equation}
 \xi(\phi) = \frac{2}{\sqrt{3}\kappa}\ln[1 + 2\alpha R] - <\Phi>
 \label{scalar and curvature}
\end{equation}
Recall that $<\Phi> =\frac{2}{\sqrt{3}\kappa} \ln[\sqrt{9 - 40\kappa^2\alpha\Lambda} - 2]$.\\
From the above expression, we can relate the boundary values of the scalar field 
(i.e $\xi(0)=v_h$ and $\xi(\pi)=v_v$) with the  Ricci scalar as,
\begin{equation}
 v_h = \frac{2}{\sqrt{3}\kappa} \ln\bigg[\frac{1 + 2\alpha R(0)}{\sqrt{9 - 40\kappa^2\alpha\Lambda} - 2}\bigg]
 \label{relation1}
\end{equation}
and 
\begin{equation}
 v_v = \frac{2}{\sqrt{3}\kappa} \ln\bigg[\frac{1 + 2\alpha R(\pi)}{\sqrt{9 - 40\kappa^2\alpha\Lambda} - 2}\bigg]
 \label{relation2}
\end{equation}
where $R(0)$ and $R(\pi)$ are the values of the curvature on Planck and TeV brane respectively. 
In section VI-A, we derive the expression of the bulk scalar curvature which in this 
scenario becomes dependent on the bulk coordinate $y$.
Thus the parameters that are used in the scalar-tensor theory  are actually related to the 
parameters of the original $F(R)$ theory.\\
Furthermore to derive the stabilization condition in scalar-tensor theory, the backreaction of the scalar field on 
spacetime geometry is neglected. It can be shown from \cite{GW1}, that this is valid as long as the stress energy tensor 
for the scalar field is less than the bulk cosmological constant which in turn implies that $v_h^2/M^3$ and $v_v^2/M^3$ are 
less than unity, where $v_h$ and $v_v$ are the boundary values of the scalar field. Now using eqn. (\ref{relation1}) 
and eqn. (\ref{relation2}), we can determine the conditions of negligible back-reactions in terms of the parameters appearing 
in the original $F(R)$ theory.\\
The effect of backreaction, though small, shall also incorporated in section VI-C. We will show that 
the backreaction modifies all the quantities described via eqn.(\ref{sol.scalar.field}, \ref{A}, \ref{B}, \ref{stabilized.modulus.ST}), 
though the modification is small 
in the limit $\kappa v_h< 1$.\\
\subsection{Radion Potential}
In this section, we consider a fluctuation of branes around the stable separation 
($r_c$). So the inter-brane separation can be considered as a field, and here, for simplicity 
we assume \cite{GW_radion} that this new field depends only on the brane coordinates. The corresponding metric ansatz is,
\begin{equation}
 d\tilde{s}^2 = e^{- 2 kT(x)|\phi|} \eta_{\mu\nu} dx^{\mu} dx^{\nu} - \tilde{T}^2(x)^2d\phi^2
 \label{grav.sol2.ST}
\end{equation}
From the perspective of four dimensional effective theory, $\tilde{T}(x)$ is known as radion field. 
Recall that the quantities in tilde are reserved for ST theory.
In order to find the radion mass, here we adopt the method proposed by Goldberger and Wise \cite{GW_radion}. 
In the GW mechanism, the same bulk scalar field which stabilize the modulus can also generate the potential 
for radion field and in the present ST theory (eqn.\ref{action2ST}), $\xi(x,\phi)$ can fulfill the purpose.

With the metric in eqn.(\ref{grav.sol2.ST}), a Kaluza-Klein reduction for the five dimensional 
Einstein-Hilbert action reduces to four dimensional effective action as,
\begin{equation}
 S_{kin}[\tilde{T}] = \frac{12M^3}{k} \int d^4x \partial_{\mu}(e^{-k\pi \tilde{T}(x)})\partial^{\mu}(e^{-k\pi \tilde{T}(x)})
 \nonumber\\
\end{equation}
As we see that $T(x)$ is not canonical and thus we redefine the field by the following transformation,
\begin{equation}
 \tilde{T}(x) \longrightarrow \tilde{\Psi}(x) = \sqrt{\frac{24M^3}{k}}e^{-k\pi \tilde{T}(x)}
 \label{trans.ST}
\end{equation}
In terms of $\tilde{\Psi}$, 
the kinetic part of radion field becomes
\begin{equation}
 S_{kin}[\tilde{\Psi}] = \frac{1}{2} \int d^4x (\partial_{\mu}\tilde{\Psi}) (\partial^{\mu}\tilde{\Psi})
 \nonumber\\
\end{equation}

Correspondingly the radion potential is obtained from eqn.(\ref{mod.potential}) by replacing $r_c$ by $\tilde{T}(x)$ i.e.
\begin{eqnarray}
 V_{eff}(\tilde{T})&=&k(\nu + 2)A^2(e^{2\nu k\tilde{T}(x)\pi} - 1)\nonumber\\ 
 &+&k(\nu - 2)B^2(1 - e^{-2\nu k\tilde{T}(x)\pi})
 \label{radion.potential.ST}
\end{eqnarray}
where $A$ and $B$ are given by,
\begin{eqnarray}
 A = v_v e^{-(2+\nu)k\tilde{T}(x)\pi} - v_h e^{-2\nu k\tilde{T}(x)\pi}\nonumber\\
 B = v_h (1 + e^{-2\nu k\tilde{T}(x)\pi}) - v_v e^{-(2+\nu)k\tilde{T}(x)\pi}
 \nonumber\\
\end{eqnarray}

Using the above potential in eqn.(\ref{radion.potential.ST}) with the transformation given in eqn.(\ref{trans.ST}), one 
lands up with the following mass squared of radion field \cite{GW_radion} in scalar-tensor theory as
\begin{equation}
 \tilde{m}^2_{rad}(ST) = \frac{k^2v_v^2}{3M^3}\epsilon^2 e^{-2kr_c\pi}
 \label{radion.mass.ST}
\end{equation}
with $\epsilon = m_{\Phi}^2/4k^2$.

\subsection{Coupling between radion and Standard Model fields}
Being a gravitational degree of freedom, radion field interacts with 
brane energy-momentum tensor and the couplings of interaction are 
constrained by four dimensional general covariance. Following 
\cite{GW_radion}, it can be shown that the coupling between radion 
and SM fields (Higgs for example) is given by
\begin{equation}
   \tilde{\lambda}_{(H-\tilde{\Psi})} = \mu^2 \sqrt{\frac{12M^3}{k}}\exp{(k\pi r_c)}
   \label{coupling 1ST}
  \end{equation}
  where $\mu$ is mass of Higgs particle and $k\pi r_c$ is obtained 
  from eqn.(\ref{stabilized.modulus.ST}). Similarly the coupling between 
  radion and $Z$ boson is,
  \begin{equation}
   \tilde{\lambda}_{(Z-\tilde{\Psi})} = m^2_Z \sqrt{\frac{12M^3}{k}}\exp{(k\pi r_c)}
   \label{coupling 2ST}
  \end{equation}
  where $m_Z$ is the mass of $Z$ boson.\\

Now we turn our focus on modulus stabilization as well as on radion mass and coupling for 
the original $F(R)$ model (eqn.(\ref{actionF(R)})) 
by using the stabilization condition of the corresponding scalar-tensor theory.

\section{Modulus stabilization, radion mass and coupling in F(R) model}

\subsection{Modulus stabilization}
Recall that the original higher curvature $F(R)$ model is presented by action given in eqn.(\ref{actionF(R)}). 
Solutions of metric ($G_{MN}$) for this $F(R)$ model can be extracted from the solutions of corresponding scalar-tensor 
theory (eqn.(\ref{grav.sol1.ST}) and eqn.(\ref{sol.scalar.field})) with the help of eqn.(\ref{conformal}). Thus 
the line element in $F(R)$ model turns out to be
\begin{equation}
 ds^2 = e^{-\frac{\kappa}{\sqrt{3}}\Phi(\phi)} [e^{- 2 kr_c|\phi|} \eta_{\mu\nu} dx^{\mu} dx^{\nu} - r_c^2d\phi^2]
 \label{grav.sol1.F(R)}
\end{equation}
where $\Phi(\phi) = <\Phi> + \xi(\phi)$ and $\xi(\phi)$ is given by eqn.(\ref{sol.scalar.field}). This solution 
of $G_{MN}$ immediately leads to the separation between hidden ($\phi=0$) and visible ($\phi=\pi$) branes along 
the path of constant $x^{\mu}$ as follows :
\begin{equation}
 \pi d = r_c \int_{0}^{\pi} d\phi e^{-\frac{\kappa}{2\sqrt{3}}\Phi(\phi)} 
 \nonumber\\
\end{equation}
where $d$ is the inter-brane separation in $F(R)$ model. 
Using the explicit functional form of $\Phi(\phi)$ (eqn.(\ref{sol.scalar.field})), above equation can be integrated 
and simplified to the following one,
\begin{equation}
 k\pi d = k\pi r_c - \frac{4k^2}{m_{\Phi}^2}\frac{\kappa v_v}{2\sqrt{3}}[\frac{v_h}{v_v} - 1]
 \label{brane.separation.F(R)}
\end{equation}
where the sub-leading terms of $\kappa\xi$ are neglected. $r_c$ is the modulus in the corresponding ST theory 
and it is stabilized which is shown in the previous section 
(eqn.(\ref{stabilized.modulus.ST})). So, it can be argued that due to the stabilization of ST theory, the modulus $d$ 
in $F(R)$ model is also stabilized with a value,
\begin{equation}
 k\pi d = \frac{4k^2}{m_{\Phi}^2}[\ln{(\frac{v_h}{v_v})} - \frac{\kappa v_v}{2\sqrt{3}}(\frac{v_h}{v_v} - 1)]
 \label{stabilized.modulus.F(R)}
\end{equation}
Hence it can be concluded that $F(R)$ model where the only independent field is spacetime 
metric ($G_{MN}$), is a 
self stabilizing system. From the expression of $m_{\Phi}^2$ (eqn.(\ref{mass_phi})), it is clear that $d$ goes to zero 
at the limit $\alpha \rightarrow 0$. Moreover for $\alpha > 0$, $m_{\Phi}^2$ becomes negative 
which in turn makes the modulus $d$ negative (see eqn.(\ref{stabilized.modulus.F(R)})), an 
unphysical situation. From the above two statements, 
it is clear that the self stabilization in $F(R)$ model arises entirely due 
to the presence of higher curvature term $\alpha R^2$ only when $\alpha < 0$.

\begin{figure}[!h]
\begin{center}
 \centering
 \includegraphics[width=3.5in,height=2.0in]{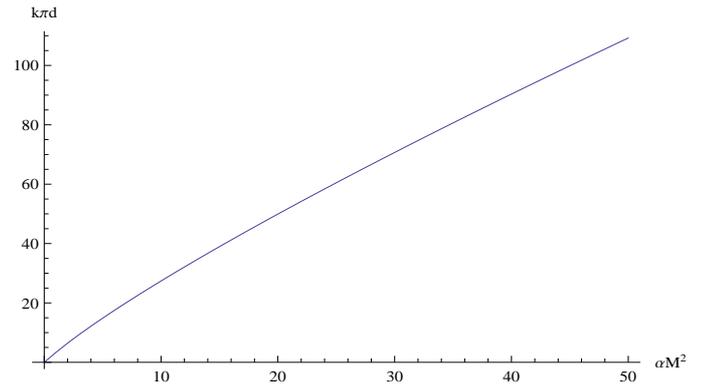}
 \caption{$k\pi d$ vs $|\alpha| M^2$}
 \label{plot alpha vs brane separation}
\end{center}
\end{figure}

From the above relation (eqn.(\ref{stabilized.modulus.F(R)})) and 
using the expression of $m_{\Phi}^2$ (eqn.(\ref{mass_phi})), we obtain the 
figure \ref{plot alpha vs brane separation} between $k\pi d$ and $|\alpha| M^2$.

The figure demonstrates that the 
brane separation ($d$) increases with the higher curvature parameter $\alpha$.\\
However, using the metric solution given in eqn.(\ref{grav.sol1.F(R)}), one calculates the five dimensional Ricci scalar as follows:
\begin{eqnarray}
 R(y)&=&-20k^2 - \frac{20}{\sqrt{3}}e^{[-\frac{\sqrt{3}}{2}\kappa cr_cy]}k^2\kappa c\nonumber\\ 
 &-&5e^{[-2\frac{\sqrt{3}}{2}\kappa cr_cy]}k^2\kappa^2 c^2
 \label{ricci_scalar_F(R)}
\end{eqnarray}
where $c$ is an integration constant. Recall that the boundary values of the curvature (i.e. $R(0)$ and $R(\pi)$, obtained from the 
eqn.(\ref{ricci_scalar_F(R)})) are related with $v_h$ and $v_v$ by eqn.(\ref{relation1}) and eqn.(\ref{relation2}) respectively. 
Thus one can tune $R(0)$, $R(\pi)$ to fix $v_h$ and $v_v$ in a desirable value.

\subsection{Radion Potential}
A fluctuation of branes around the stable configuration $d$ is now considered. This fluctuation 
can be taken as a field ($T(x)$) and for simplicity, this new field is assumed to be the 
function of brane coordinates only. The metric takes the following form,
\begin{eqnarray}
 ds^2 = e^{-\frac{\kappa}{\sqrt{3}}\Phi(x,\phi)} &[&e^{- 2 kT(x)|\phi|} 
 \eta_{\mu\nu} dx^{\mu} dx^{\nu}\nonumber\\ 
 &-&T(x)^2d\phi^2]
 \label{grav.sol2.F(R)}
\end{eqnarray}
From the angle of four dimensional effective theory, $T(x)$ is known as radion field. Moreover 
$\Phi(x,\phi)$ is obtained from eqn.(\ref{sol.scalar.field}) by replacing $r_c$ to $T(x)$. 
Using the metric in eqn.(\ref{grav.sol2.F(R)}), a Kaluza-Klein reduction for five dimensional Einstein-Hilbert 
action generates a kinetic as well as a potential part for the radion field $T(x)$. Kinetic part 
comes as
\begin{eqnarray}
 S_{kin}[T]&=&\frac{12M^3}{k} (1 - \frac{40}{\sqrt{3}}\alpha k^2\kappa v_h)\nonumber\\
 &\int&d^4x \partial_{\mu}(e^{-k\pi T(x)})\partial^{\mu}(e^{-k\pi T(x)}) 
 \nonumber\\
\end{eqnarray}
It is evident that $T(x)$ is not canonical and in order to make it canonical, we 
redefine the field as 
\begin{equation}
 T(x) \longrightarrow \Psi(x) = f e^{-k\pi T(x)}
 \label{trans.F(R)}
\end{equation}
where $f = \sqrt{\frac{24M^3}{k}} [1 - \frac{20}{\sqrt{3}}\alpha k^2\kappa v_h]$. 
For $\alpha \rightarrow 0$, the action contains only the linear term in Ricci scalar 
and the factor $f$ goes to $\sqrt{24M^3/k}$ which agrees with \cite{GW_radion}.
With the redefined field, kinetic part of radion field becomes,
\begin{equation}
 S_{kin}[\Psi] = \frac{1}{2} \int d^4x (\partial_{\mu}\Psi) (\partial^{\mu}\Psi)
 \nonumber\\
\end{equation}
Finally the Kaluza-Klein reduction leads to the potential part of radion field as follows,
\begin{eqnarray}
 V(\Psi)&=&-\frac{20}{\sqrt{3}}\frac{\alpha k^5}{M^6} \Psi^4 
 [v_v - (v_h - \frac{\kappa v_h^2}{2\sqrt{3}}\nonumber\\ 
 &+&\frac{\kappa v_hv_v}{2\sqrt{3}})(\Psi/f)^\epsilon ]^2
 \label{potential.radion.F(R)}
\end{eqnarray}
It may be observed that $V(\Psi)$ goes to zero as $\alpha$ tends 
to zero. This is expected because for $\alpha \rightarrow 0$, 
the action contains only the Einstein part which does not produce any 
potential term for the radion field \cite{GW_radion}. 
Thus for five dimensional warped geometric model, the radion potential is 
generated from the higher order curvature term $\alpha R^2$. 
The potential in eqn.(\ref{potential.radion.F(R)}) has a vev at
\begin{equation}
 <\Psi> = f (\frac{v_v}{v_h})^{1/\epsilon} [1 - \frac{\kappa v_h}{2\sqrt{3}}(1 - \frac{v_v}{v_h})]^{-1/\epsilon}
 \label{vev.radion.F(R)}
\end{equation}
which leads to the inter-brane separation as,
\begin{eqnarray}
 k\pi d&=&k\pi <T_{F(R)}(x)>\nonumber\\ 
 &=&\frac{4k^2}{m_{\Phi}^2}[\ln{(\frac{v_h}{v_v})} - \frac{\kappa v_h}{2\sqrt{3}}(1 - \frac{v_v}{v_h})]
 \label{modulus}
\end{eqnarray}
in the leading order of $\kappa\xi$. This above equation exactly resembles with the eqn.(\ref{stabilized.modulus.F(R)}) 
which again indicates that the five dimensional warped spacetime we considered, is self stabilized 
by higher curvature gravitational theory. We emphasize that due to the presence 
of conformal factor connecting the two theories, the value of 
$k\pi d$ (in $F(R)$ model) is less than $k\pi r_c$ (in ST model). However 
we find that the stabilization of modulus remain intact in both the 
models. Finally 
the squared mass of radion field is as follows,
\begin{eqnarray}
 m_{rad}^2(F(R))&=&-\frac{20}{\sqrt{3}}\frac{\alpha k^4}{M^6} \epsilon^2 e^{-2kd\pi} v_h^2v_v^2\nonumber\\
 &[&1 - \frac{40}{\sqrt{3}}\alpha k^2\kappa v_h] [\frac{v_h}{v_v} - 1]^2
 \label{radion.mass.F(R)}
\end{eqnarray}
It is noticed that the mass of radion field is enhanced by the higher curvature 
terms in five dimensional gravitational action.\\

\subsection{Radion potential with backreaction}
It may be mentioned that the stabilized interbrane separation obtained in eqn. (\ref{modulus}) is based on 
the conditions that the bulk scalar potential is retained up to quadratic term (see eqn.(\ref{action2ST})) 
and the backreaction of the bulk scalar field is neglected on five dimensional spacetime. Both these 
conditions are followed from the assumption that $\kappa v_h< 1$. Relaxation of this assumption is crucial 
to check the status of the stabilization condition in the presence of higher order self interaction 
terms in the bulk scalar potential. Here we examine whether the radion potential admits any stability 
when $V(\Phi)$ is retained up to cubic term in $\Phi$. In this scenario, the five dimensional action 
in ST theory turns out to be,
\begin{eqnarray}
 S&=&\int d^4x d\phi \sqrt{\tilde{G}} \bigg[-\frac{\tilde{R}}{2\kappa^2} + \frac{1}{2}\tilde{G}^{MN}\partial_M\xi 
 \partial_N\xi\nonumber\\
 &-&(1/2)m_{\Phi}^2\xi^2 - \frac{g}{3}\xi^3 - \Lambda_{eff}\bigg]
 \label{action2ST_final}
\end{eqnarray}
where $g$ is the self cubic coupling of $\Phi(\phi)$ and can be easily determined from the form 
of $V(\Phi)$ presented in eqn.(\ref{scalar_potential}) as,
\begin{eqnarray}
 g = \frac{\sqrt{3}\kappa}{16\alpha}\frac{[2\sqrt{9 - 40\kappa^2\alpha\Lambda} + 3]}{[\sqrt{9 - 40\kappa^2\alpha\Lambda} - 2]^{\frac{2}{3}}}
 \label{g}
\end{eqnarray}
Considering the metric ansatz as,
\begin{equation}
 d\tilde{s}^2 = e^{- 2 A(\phi)} \eta_{\mu\nu} dx^{\mu} dx^{\nu} - r_c^2d\phi^2
 \label{grav.sol1.ST_final}
\end{equation}
the gravitational as well as the scalar field equations of motion take the following form,
\begin{eqnarray}
 \frac{4}{r_c^2} A'^2(\phi) - \frac{1}{r_c^2}A''(\phi) = -(2\kappa^2/3)(\frac{1}{2}m_{\Phi}^2\xi^2 + \frac{g}{3}\xi^3)
 \label{equation1_final}
\end{eqnarray}

\begin{eqnarray}
  \frac{1}{r_c^2}A'^2(\phi)= \frac{\kappa^2}{12r_c^2}\xi'^2 - (\kappa^2/6)(\frac{1}{2}m_{\Phi}^2\xi^2 + \frac{g}{3}\xi^3)
 \label{equation2_final}
\end{eqnarray}

\begin{eqnarray}
 \frac{1}{r_c^2}\xi''(\phi)=\frac{4}{r_c^2}A'\xi' + m_{\Phi}^2\xi(\phi) + g\xi^2(\phi)
 \label{equation3_final}
\end{eqnarray}
To determine the solutions of the above differential equations, we apply the iterative method 
by considering the form of metric determined in eqn. (\ref{grav.sol1.ST}) as the zeroth order solution. 
In the leading order correction of $\kappa v_h$, $\xi(\phi)$ and $A(\phi)$ turn out to be
\begin{eqnarray}
 \xi(\phi)&=&\big[Ae^{(2+\nu)kr_c\phi} + Be^{(2-\nu)kr_c\phi}\big]\nonumber\\
 &+&\frac{\kappa v_h}{16\sqrt{3}\alpha k^2} \frac{[2\sqrt{9 - 40\kappa^2\alpha\Lambda} + 3]}
 {[\sqrt{9 - 40\kappa^2\alpha\Lambda} - 2]^{\frac{2}{3}}}*\nonumber\\
 &\bigg[&\frac{\exp{\big[2(2+\nu)kr_c\phi-4\nu kr_c\pi\big]}}{m_{\Phi}^2r_c^2 + 8kr_c(2+\nu)-4(2+\nu)^2}\nonumber\\
 &+&\frac{2v_v}{v_h}\frac{\exp{\big[4kr_c\phi-(2+\nu)kr_c\pi\big]}}{m_{\Phi}^2r_c^2 + 16kr_c- 16}\bigg]
 \label{sol_scalar_final}
\end{eqnarray}
and

\begin{eqnarray}
 A(\phi)&=&kr_c\phi + \frac{\kappa^2v_h^2}{12}\bigg[e^{-4\nu kr_c\pi} e^{-2(2+\nu)kr_c\phi}\nonumber\\
 &+&e^{2(2-\nu)kr_c\phi} \big(1+\frac{v_v}{v_h}e^{-(2+\nu)kr_c\pi}\big)\bigg]
 \label{sol_warp_final}
\end{eqnarray}
Thus due to the inclusion of the bulk scalar field backreaction, the warp factor gets modified and 
the correction term is proportional to $\kappa^2v_h^2$ which is indeed small for $\kappa v_h< 1$. However 
if one includes these corrections in $\xi(\phi)$ and $A(\phi)$, one can extract the modified solution in F(R) model 
by a conformal transformation as indicated earlier:
\begin{eqnarray}
 ds^2&=&\exp{\big[-\frac{\kappa}{\sqrt{3}}(<\Phi>+\xi(\phi))\big]}*\nonumber\\ 
 &[&e^{- 2 A(\phi)} \eta_{\mu\nu} dx^{\mu} dx^{\nu} - r_c^2d\phi^2]
 \label{grav.sol1.F(R)_final}
\end{eqnarray}
where $ds^2$ is the line element in $F(R)$ model and $\xi(\phi)$, $A(\phi)$ are 
given in eqn.(\ref{sol_scalar_final}) and eqn.(\ref{sol_warp_final}) respectively. Plugging back 
the above solution of metric (in eqn.(\ref{grav.sol1.F(R)_final})) into the five dimensional 
$F(R)$ action and integrating over $\phi$, the radion potential is obtained as
\begin{eqnarray}
 V(\Psi)&=&-\frac{20}{\sqrt{3}}\frac{\alpha k^5}{M^6} \Psi^4 
 \bigg[v_v - \big(v_h - \frac{\kappa v_h^2}{2\sqrt{3}} + \frac{\kappa v_hv_v}{2\sqrt{3}}\nonumber\\
 &+&\frac{\kappa^2v_h^3}{64\sqrt{3}}-\frac{\kappa^2v_h^2v_v}{616\sqrt{3}} + \frac{\kappa^2v_v^2v_h}{16\sqrt{3}}\big)(\Psi/f)^\epsilon\bigg]^2
 \label{potential.radion.F(R)_final}
 \end{eqnarray}
 The above potential has a stable minimum at,
 \begin{eqnarray}
 <\Psi>&=&f (\frac{v_v}{v_h})^{1/\epsilon} \bigg[1 - \frac{\kappa v_h}{2\sqrt{3}}(1 - \frac{v_v}{v_h})+\frac{\kappa^2v_h^2}{64\sqrt{3}}\nonumber\\
 &-&\frac{\kappa^2v_hv_v}{16\sqrt{3}} + \frac{\kappa^2v_v^2}{16\sqrt{3}}\bigg]^{-1/\epsilon}
 \label{vev.radion.F(R)_final}
\end{eqnarray}
which leads to the stabilized interbrane separation as:
\begin{eqnarray}
 k\pi d&=&k\pi <T_{F(R)}(x)>\nonumber\\ 
 &=&\frac{4k^2}{m_{\Phi}^2}\bigg[\ln{(\frac{v_h}{v_v})} - \frac{\kappa v_h}{2\sqrt{3}}(1 - \frac{v_v}{v_h})\nonumber\\
 &+&\frac{\kappa^2v_h^2}{64\sqrt{3}}(1-4\frac{v_v}{v_h}+ 4\frac{v_v^2}{v_h^2})\bigg]
 \label{modulus_final}
\end{eqnarray}

Comparing eqn.(\ref{modulus}) and eqn.(\ref{modulus_final}), it can be seen that the vev of the radion 
field and hence the stable modulus is shifted due to the presence of 
higher order self interaction terms in the bulk scalar potential or the inclusion of scalar field backreaction. However 
this shift is indeed small in the limit $\kappa v_h< 1$\\

\subsection{Coupling between radion and Standard Model fields}\label{radion_coupling}
The radion field arises as a scalar degree
of freedom on the TeV brane and has interactions  with the Standard Model (SM) 
fields. 
  From the five dimensional metric (eqn.(\ref{grav.sol2.F(R)})), 
it is clear that the induced metric on visible brane is 
  $(\frac{\Psi}{f})^2e^{-\kappa v_v\frac{1}{\sqrt{3}}}\eta_{\mu\nu}$ 
  (where $f = \sqrt{\frac{24M^3}{k}} [1 - \frac{20}{\sqrt{3}}\alpha k^2\kappa v_h]$) 
and consequently $\Psi(x)$ couples directly with SM fields. \\
For example, consider the 
  Higgs sector of Standard Model,
  \begin{eqnarray}
   S_{Higgs}&=&(1/2) \int d^4x (\Psi/f)^4 e^{-\kappa v_v\frac{2}{\sqrt{3}}}\nonumber\\ 
&[&(\Psi/f)^{-2} e^{\kappa v_v\frac{1}{\sqrt{3}}}\eta^{\mu\nu}\partial_\mu h\partial_\nu h - \mu_0^2h^2]
   \nonumber\\
  \end{eqnarray}
  where $h(x)$ is the Higgs field. In order to get a canonical
 kinetic term, one needs to redefine $h(x) \longrightarrow H(x) = \frac{<\Psi>}{f}h(x)$. 
Therefore for $H(x)$, the above action can be written as,
  \begin{eqnarray}
   S_{Higgs}&=&(1/2) \int d^4x [(\frac{\Psi}{<\Psi>})^{2}\eta^{\mu\nu}\partial_\mu H\partial_\nu H\nonumber\\
&-&(\frac{\Psi}{<\Psi>})^{4}\mu^2H^2]
   \label{higgs action}
  \end{eqnarray}
  where $\mu = \mu_0 \frac{<\Psi>}{f}e^{-\kappa v_v\frac{1}{\sqrt{3}}}$. 
Considering a fluctuation of $\Psi(x)$ about its vev as 
  $\Psi(x)=<\Psi> + \delta\Psi$, one can obtain (from eqn.(\ref{higgs action})) 
that $\delta\Psi$ couples to $H(x)$ through the trace of the 
  energy-momentum tensor of the Higgs field:
  \begin{equation}
   \mathcal{L} = \frac{\delta\Psi}{<\Psi>}T^\mu_\mu(H)
   \nonumber\\
  \end{equation}
  So, the coupling between radion and Higgs field become, 
 $\lambda_{(H-\delta\Psi)} = \frac{\mu^2}{<\Psi>}$. Similar consideration holds for any other 
  SM fields. For example for $Z$ boson, the corresponding coupling is 
$\lambda_{(Z-\delta\Psi)} = \frac{m^2_Z}{<\Psi>}$. Thus the inverse of $<\Psi>$ plays 
  a crucial role in determining the coupling strength
 between radion and SM fields. In the present case, we obtain
  \begin{equation}
  <\Psi> = f e^{-k\pi d}
   \nonumber\\
  \end{equation}
  Hence finally we arrive at,
  \begin{eqnarray}
   \lambda_{(H-\delta\Psi)}&=&\mu^2 \sqrt{\frac{k}{24M^3}} e^{k\pi d}\nonumber\\
   &*&[1 + \frac{20}{\sqrt{3}}\alpha k^2\kappa v_h]
   \label{coupling 1}
  \end{eqnarray}
  and
  \begin{eqnarray}
   \lambda_{(Z-\delta\Psi)}&=&m^2_Z \sqrt{\frac{k}{24M^3}} e^{k\pi d}\nonumber\\
   &*&[1 + \frac{20}{\sqrt{3}}\alpha k^2\kappa v_h]
   \label{coupling 2}
  \end{eqnarray}
in the leading order of $\kappa$. \\
The coupling between radion and fermion field is similarly given by,
\begin{eqnarray}
   \lambda_{(fermion-\delta\Psi)}&=& \sqrt{\frac{k}{24M^3}} e^{k\pi d}\bigg[1 + \frac{20}{\sqrt{3}}\alpha k^2\kappa v_h\bigg]
   \label{coupling 3}\\
  \end{eqnarray}
  
  If the fermion fields are allowed to propagate in the bulk, then the coupling with radion changes 
  from that given in  (eqn.(\ref{coupling 3})). This can be determined
  by  Kaluza-Klein (KK) mode expansion of the bulk fermion \cite{grossman} in a bulk governed by $F(R)$  gravity. Here we focus  
  on the zeroth order KK mode and  the solution of the zeroth order  KK mode wave function in the bulk 
  takes the form: 
  \begin{equation}
   \chi_L(\phi) = [\sqrt{9 - 40\kappa^2\alpha\Lambda} - 2]^{-\frac{1}{3}} \sqrt{\frac{(k+2m)d}{e^{(k+2m)\pi d} -1}} e^{md\phi}
   \nonumber
  \end{equation}
  where $\chi_L(\phi)$ is the left handed fermion wave function and $m$ is bulk fermionic mass. Recall that 
  $d$ is the interbrane separation and $<\Phi>$ is given in eqn. (\ref{vev_phi}). Similarly for the right handed mode ($\chi_R(\phi)$),
  \begin{equation}
   \chi_R(\phi) = [\sqrt{9 - 40\kappa^2\alpha\Lambda} - 2]^{-\frac{1}{3}} \sqrt{\frac{(k-2m)d}{e^{(k-2m)\pi d} -1}} e^{-md\phi}
   \nonumber\\
  \end{equation}
  Using these above solutions, we determine the coupling of radion with zeroth order fermionic KK mode, which yields,
\begin{eqnarray}
   \lambda_L&=&\sqrt{\frac{k}{24M^3}} e^{k\pi d}\bigg[1 + \frac{20}{\sqrt{3}}\alpha k^2\kappa v_h\bigg]\nonumber\\
   &[&\sqrt{9 - 40\kappa^2\alpha\Lambda} - 2]^{-\frac{2}{3}} (k+2m)d\nonumber\\
   &\bigg(&\frac{e^{(k+2m)\pi d}}{e^{(k+2m)\pi d} -1}\bigg)
   \label{coupling 4}
  \end{eqnarray}
  for left handed chiral mode and,
  \begin{eqnarray}
  \lambda_R&=&\sqrt{\frac{k}{24M^3}} e^{k\pi d}\bigg[1 + \frac{20}{\sqrt{3}}\alpha k^2\kappa v_h\bigg]\nonumber\\
   &[&\sqrt{9 - 40\kappa^2\alpha\Lambda} - 2]^{-\frac{2}{3}} (k-2m)d\nonumber\\
   &\bigg(&\frac{e^{(k-2m)\pi d}}{e^{(k-2m)\pi d} -1}\bigg)
   \label{coupling 5}
  \end{eqnarray}
  for right handed mode.\\
  The form of $v_h$ and $v_v$ in terms of five dimensional Ricci scalar can be extracted from 
  eqn. (\ref{relation1}) and  eqn. (\ref{relation2}). 
  It is evident that the coupling $\lambda_L$, $\lambda_R$ is modified  by the factor $(k\pm 2m)d$, 
  in comparison to the coupling given in eqn. (\ref{coupling 3}) for  the fermion fields 
  confined on the TeV brane.\\
   From  above analysis  we note that the coupling between radion and SM fields 
is suppressed due to the presence of higher curvature parameter $\alpha$ which in turn modifies the phenomenology on visible 3-brane.\\
Before concluding, we mention about a recent work \cite{ssg3} ,  where a higher curvature gravity model with $R^4$ terms in the action is 
considered. The corresponding conformally transformed scalar action includes a quartic term which resembles closely to the scalar action 
considered in reference [11 and 13]. 
It has been shown that with such a specific choice of the scalar action one can estimate the exact modification of the 
warp factor due to the effects of the back-reaction of the scalar field on the background geometry and thus it enables 
us to address the role of back-reaction on the stabilization issue and  also on various parameters of the low-energy 
effective action. However in such a  model,  the back-reacted geometry can be exactly determined if the scalar mass 
and the quartic term in the potential  are inter-related and in the limit of the  quartic term going to zero, 
the mass term also goes to zero. Therefore there is no smooth limit which takes this model   to that considered by the  
GW where  only quadratic mass term was present.
The  work reported in this article  however  has a different goal from that of \cite{ssg3}. Here we show that in the 
leading order quadratic curvature correction to Einstein action in the bulk we find a dual scalar tensor theory which 
under certain approximation is similar to the original GW scalar action which has  a quadratic mass term only.  
We therefore explore and re-examine the originally proposed  Goldberger-Wise modulus stabilization condition  
in the light of higher curvature  gravity models where such a  stabilizing field appears naturally from higher 
curvature degrees of freedom with a minimal  curvature extension.\\

\section{Conclusion}
In this work, we consider a five dimensional AdS, compactified warped geometry model with two 3-branes 
embedded within the spacetime. Due to large curvature ($\sim$ Planck scale) in the bulk, the spacetime is assumed to be governed by a higher curvature gravity  model , $F(R) = R + \alpha R^2$. In this scenario, we 
address how the modulus stabilization and radion phenomenology are affected by higher curvature term. 
The findings and implications of our results are as follows :
\begin{itemize}
 \item The model comes as a self stabilizing system due to the presence of 
 higher curvature term $\alpha R^2$.  This is in sharp contrast to a model with only Einstein term  in the bulk where the modulus can not be stabilized without incorporating any external degrees of freedom such as a scalar field. However for the higher curvature gravity model, this additional degree of freedom originates naturally 
 from the higher curvature term and plays the role of a stabilizing field. It may also be noted that for $\alpha \rightarrow 0$, the stabilization 
 condition (eqn.(\ref{stabilized.modulus.F(R)})) leads to zero  brane separation.
 
 \item We scan the parametric space of $\alpha$ for which the modulus 
 is going to be stabilized. Our result reveals that for $\alpha > 0$, the inter-brane separation 
 becomes negative which is an unphysical situation. Thus the braneworld 
 we have considered is stabilized only when $\alpha < 0$. This puts
 constraints on the $F(R)$ model itself. Moreover the distance between the branes increases with 
 the value of the parameter $\alpha$ which is evident from figure (\ref{plot alpha vs brane separation}).
 Thus the results obtained in this work clearly bring out the correlation between a geometrically stable  warped solution resulting from negative bulk curvature and the stability of the higher curvature $F(R)$ model free from ghosts.
 
 \item Quadratic term in curvature also generates the radion potential with a stable 
 minimum. We find the 
 radion mass as well as radion coupling with SM fields. The 
 expressions of mass (eqn.(\ref{radion.mass.F(R)})) and coupling 
 (eqn.(\ref{coupling 1}) 
 and eqn.(\ref{coupling 2})) clearly indicate that the radion mass is enhanced while the coupling is suppressed in 
 comparison to  the scenario where only Einstein gravity resides in the bulk \cite{GW_radion}. 
 Thus the cross section between radion and SM fields is overall suppressed due to 
 the presence of higher order curvature terms in five dimensional gravitational action 
 leading to a possible explanation of the invisibility of the radion field in the present experimental resolution.
 
 \item The effect of backreaction on the radion potential and its minima are studied. It is shown that 
 the corrections due to the backreaction is indeed small in the limit $\kappa v_h< 1$. The possible correction terms 
 for the backreaction are determined.
\end{itemize}


\begin{thebibliography}{90}
\bibitem{arkani}
N. Arkani-Hamed, S. Dimopoulos, G. Dvali, Phys. Lett.
B 429 (1998) 263; N. Arkani-Hamed, S. Dimopoulos, G.
Dvali, Phys. Rev. D 59 (1999) 086004; I. Antoniadis, N.
Arkani-Hamed, S. Dimopoulos, G. Dvali, Phys. Lett. B
436, 257 (1998)


\bibitem{horava}
P. Horava and E. Witten, Nucl. Phys. B475, 94 (1996); B460, 506 (1996)

 \bibitem{RS1}
L. Randall and R. Sundrum, Phys. Rev. Lett. {\bf 83}, 3370 (1999);

\bibitem{kaloper}
 N. Kaloper, Phys. Rev. D60, 123506 1999; T. Nihei,Phys. Lett. B465, 81 (1999); H. B. Kim and H. D. Kim,Phys. Rev. D61, 064003 (2000)
 
 \bibitem{cohen}
  A. G. Cohen and D. B. Kaplan, Phys. Lett. B470, 52(1999); 
  
  \bibitem{burgess}
  C. P. Burgess, L. E. Ibanez, and F. Quevedo,ibid. 447, 257 (1999); 
  
  \bibitem{chodos}
  A. Chodos and E. Poppitz, ibid.471, 119 (1999); T. Gherghetta and M. Shaposhnikov,Phys. Rev. Lett.85, 240 (2000)
  
  \bibitem{Green}M. B. Green, J. H. Schwarz and E. Witten, ``Superstring Theory'',Vol.I
  and Vol.II,Cambridge University Press(1987),
String Theory, J.Polchinski, Cambridge University Press(1998)

  \bibitem{GW1} W. D. Goldberger and M. B. Wise, Phys.Rev.Lett.{\bf 83}, 4922 (1999).
  
  \bibitem{GW_radion} W. D. Goldberger and M. B. Wise, Phys.Lett B 475, 275-279 (2000)
  
\bibitem{kribs} C. Csaki, M. L. Graesser and Graham D. Kribs,  Phys.
Rev.D.{\bf 63}, 065002 (2001).

\bibitem{julien}
J. Lesgourgues, L. Sorbo, Goldberger-Wise variations: Stabilizing brane models with a bulk scalar, Phys. Rev. D69, 084010 (2004)

\bibitem{wolfe} O. DeWolfe, D. Z. Freedman, S. S. Gubser and A. Karch, Phys.
Rev.D.{\bf 62}, 046008 (2000).

  \bibitem{dhr}
H. Davoudiasl, J. L. Hewett, T. G. Rizzo,Phys.Rev. Lett. 84, 2080 (2000)

\bibitem{rizzo}
T. G. Rizzo, Int.J.Mod.Phys A15, 2405-2414 (2000)

\bibitem{yong}
Y. Tang, JHEP 1208, 078 (2012)

\bibitem{dhr1}
H. Davoudiasl, J.L. Hewett, T.G. Rizzo, JHEP 0304, 001 (2003)

\bibitem{thomas}
M. T. Arun, D. Choudhury, A. Das, S, Sengupta, Phys.Lett.B746, 266-275 (2015)

\bibitem{wiseman}
P. Figueras, T. Wiseman: Gravity and Large Black Holes in Randall-Sundrum II Braneworlds, PRL 107, 081101 (2011)

\bibitem{naresh}
N. Dadhich, R. Maartens, P. Papadopoulos, V. Rezania : Black Holes on the Brane, Phys.Lett. B487 (2000)

\bibitem{dai}
D. C. Dai, D. Stojkovic : Analytic solution for a static black hole in RSII model, Phys.Lett. B704, 354-359 (2011)

\bibitem{atlas1}
ATLAS Collaboration, Phys.Lett.B710, 538-556 (2012)

\bibitem{atlas2}
ATLAS Collaboration, G. Aad et al, Phys.Rev.D.90, 052005 (2014)

 \bibitem{faraoni}
 T. P. Sotiriou and V. Faraoni, "f(R) Theories Of Gravity," Rev.Mod.Phys.82, 451–497 (2010), arXiv:0805.1726 [gr-qc]
 
 \bibitem{felice}
 A.De Felice and S. Tsujikawa, "f(R) theories," Living Rev.Rel. 13, 3 (2010), arXiv:1002.4928 [gr-qc].
 
 \bibitem{paliathanasis}
 A.Paliathanasis, "f(R)-gravity from Killing Tensors," Class. Quant. Grav.33no. 7, 075012 (2016),\\
 arXiv:1512.03239 [gr-qc].
 
 \bibitem{nojiri2}
  S.Nojiri, S. D. Odintsov, Phys.Lett.B 631 (2005) 1.arxiv:hep-th/0508049
  
  \bibitem{nojiri3}
  S.Nojiri, S. D. Odintsov, O.G.Gorbunova, J.Phys.A39, 6627 (2006). arxiv:hep-th/0510183
  
  \bibitem{cognola}
  G.Cognola, E. Elizalde, S. Nojiri, S.D. Odintsov, S.Zerbini, Phys.Rev. D73, 084007 (2006)
  
  \bibitem {kim1} J.E.Kim, B. Kyae, H.M. Lee, Phys.Rev D62, 045013 (2000), arxiv: hepph/9912344.
  
  \bibitem{kim2} J.E.Kim, B. Kyae, H.M. Lee, Nucl. Phys.B582, 296 (2000) , Erratum : Nucl.Phys.B591,587 (2000), hep-th/0004005.
  
  \bibitem{ssgsc} S. Choudhury and S. SenGupta, JHEP 1302, 136 (2013)
  
\bibitem{barrow}
 J. D. Barrow and S. Cotsakis, "Inflation and the Conformal Structure of Higher Order Gravity Theories,"\\
 Phys. Lett.B214, 515†- 518 (1988).
  
  \bibitem{marino}
  S. Capozziello, R. de Ritis, and A. A. Marino, "Some aspects of the cosmological conformal equivalence\\ 
  between 'Jordan frame' and 'Einstein frame',\\ 
  "Class. Quant. Grav.14, 3243–3258 (1997), arXiv: gr-qc/9612053 [gr-qc].
  
  \bibitem{bahamonde}
  S. Bahamonde, S. D. Odintsov, V. K. Oikonomou, and M. Wright, "Correspondence of F(R) Gravity Singularities\\ 
  in Jordan and Einstein Frames, "arXiv: 1603.05113 [gr-qc].
  
  \bibitem{catena}
  R. Catena, M. Pietroni, and L. Scarabello, "Einstein and Jordan reconciled:\\
  a frame-invariant approach to scalar-tensor cosmology, "Phys. Rev.D76, 084039 (2007),\\
  arXiv: astro-ph/0604492 [astro-ph].
  
  \bibitem{ssg1}
  S. SenGupta, S. Chakraborty, ``Solving higher curvature gravity theories'',  Eur.Phys.J. C76, no.10, 552 (2016)\\
  arXiv: 1604.05301.
  
  \bibitem{ssg2}
  S. Anand, D. Choudhury, Anjan A. Sen, S. SenGupta, "A Geometric Approach to Modulus Stabilization"\\
   Phys.Rev. D92, no.2, 026008 (2015), arXiv:1411.5120.
   
  \bibitem{ssg3}  S. Chakraborty and S. SenGupta, arXiv:1701.01032

\bibitem{grossman} Y Grossman and M. Neubert,   Phys.Lett.B474, 361-371 ( 2000)

\end{thebibliography}
\end{document}